# Software Analytics to Software Practice: A Systematic Literature Review

Tamer Mohamed Abdellatif, Luiz Fernando Capretz, and Danny Ho
Department of Electrical & Computer Engineering
Western University
London, Ontario, Canada
tmohame7@uwo.ca, lcapretz@uwo.ca, danny@nfa-estimation.com

*Abstract*—Software Analytics (SA) is a new branch of big data analytics that has recently emerged (2011). What distinguishes SA from direct software analysis is that it links data mined from many different software artifacts to obtain valuable insights. These insights are useful for the decision-making process throughout the different phases of the software lifecycle. Since SA is currently a hot and promising topic, we have conducted a systematic literature review, presented in this paper, to identify gaps in knowledge and open research areas in SA. Because many researchers are still confused about the true potential of SA, we had to filter out available research papers to obtain the most SA-relevant work for our review. This filtration yielded 19 studies out of 135. We have based our systematic review on four main factors: which software practitioners SA targets, which domains are covered by SA, which artifacts are extracted by SA, and whether these artifacts are linked or not. The results of our review have shown that much of the available SA research only serves the needs of developers. Also, much of the available research uses only one artifact which, in turn, means fewer links between artifacts and fewer insights. This shows that the available SA research work is still embryonic leaving plenty of room for future research in the SA field.

## I. INTRODUCTION

Software analytics (SA) represents a branch of big data analytics. SA is concerned with the analysis of all software artifacts, not only source code. Its importance comes from the need to extract support insights and facts from the available software artifacts to facilitate decision making. Artifacts are available from all software development life cycle steps, beginning with the proposal and project initiation phases and ending with the project closure and customer satisfaction surveys. The dynamic nature of the software industry is associated with decision-making needs through all software business tiers. These tiers vary from the higher level of the management board and setting the enterprise vision and portfolio management, going through project management planning and implementation by software developers. As emphasized by some experts [1-4] in the SA domain, all of the stakeholders involved deserve to be supported with decision-making tools in order to facilitate the decision-making process. SA can play the role of the tool provider by providing suitable and supportive insights and facts to software industry stakeholders to make their decision making easier, faster, more precise, and more confident. The main difference between SA and direct software analysis is that rather than just providing straightforward insights extraction SA performs additional advanced steps. As clarified by Hassan [1], SA must provide visualization and useful interpretation of insights in order to facilitate decision making.

This paper is organized as follows: In Section II, we will illustrate our review methodology. Our review results are illustrated in Section III. Section IV presents the limitation of this review and, finally, we conclude our work in Section V.

## II. METHOD

In systematic literature review (SLR), we followed Kitchenham [5] approach for a software engineering literature review. So we started with the planning phase in which we developed the review protocol. We conducted our review in six stages: defining research questions, designing the search strategy, selecting the studies to use, assessing the quality, extracting data, and synthesizing data.

*A. Research Questions*

In our SLR, we are trying to answer the following research questions:

1. **RQ1: Which software practitioners does the available SA research target?**

RQ1 aims to identify the beneficiary stakeholders from available SA studies. It also aims to assess whether SA studies target different levels of stakeholders or only focus on the software development team in order to draw the attention of the SA research community to improve the research plan.

2. **RQ2: Which domains are covered by SA studies?**

RQ2 tries to highlight the scope of the available SA studies. The target domains, such as software maintainability and incident management, will be determined. Practitioners can interpret this information from two points of view. The first point of view is to know SA hot topics and consider these for their research plan, while the other view is to analyze any research gap and take the lead to consider this as an original research point.

*3. RQ3: Which software artifacts are extracted?*

The main difference between SA and direct software analysis is making use of all of the available artifacts in order to come out with strong decision support insights. Therefore, RQ3 aims to verify that this idea is clear for the current research community.

*4. RQ4: If different artifacts are used, are they linked together?*

RQ4 tries to evaluate if each study satisfies the SA main idea of linking different software artifacts. This linkage aims to come out with more advanced insights unlike direct software analysis and metrics where researchers use each artifact separately without linkage to other artifacts.

*B. Search Strategy*

*1) Search Terms*

To guarantee that the review is closely relevant to SA, we tried to limit our search to the most SA relevant search term. So, we started with the term "software analytics" then we went through the following steps:
1. Extracting the major distinct terms from our research questions.
2. Using different spellings of the terms.
3. Updating our search term with keywords from relevant papers.
4. Using the main alternatives and adding "OR operator" in order to get the maximum amount of directly relevant literature.

These steps yielded the following search term:
"Software analytics" OR "Software analytic" OR "Software development analytics" OR "Software development analytic".

*2) Literature Resources*

We included two electronic databases in order to search for our primary review studies: *IEEE – Xplore* and *ACM Digital Library*. The search term was constructed using the advanced search features provided by each of these two databases. The search covered metadata in the case of *IEEE – Xplore*, and both metadata and body (content) of literatures in the case of *ACM Digital Library*.

Our search included the period January 2000 to December 2014. As SA concept was initially introduced by Zhang et al. in 2011 [6], we expected that relevant literatures would be from 2011 and forward, but we made our search timeframe wider in order to guarantee gathering all of the relevant papers.

*C. Study Selection*

The search results contained 135 unique candidate papers (41 papers from IEEE Xplore, 102 from ACM Digital Library and 8 duplicate papers between the two databases which were removed). In order to eliminate any irrelevant papers which would not add any significant information, we conducted the following two filtration phases:
- Filtration phase 1: both inclusion and exclusion criteria were defined and applied to the unique candidate papers to eliminate any irrelevant papers so that only relevant papers with useful information would result from this phase.
- Filtration phase 2: the quality assessment criteria (as defined in the next section) were used to assess candidate papers that are output from phase 1. The papers which satisfy the quality boundary will be used in the data extraction stage.

The following were the inclusion and exclusion criteria:

**Inclusion criteria**
- SA concepts were applied to extract insights from software project artifacts.
- Research was relevant to software project lifecycle phases.
- Research was directly related to the software industry and stakeholders.
- For duplicate publications of the same study, the newest and most complete one was selected. This is recorded for only one study whose related work appeared in two conferences.

**Exclusion criteria**
- Studies that were irrelevant to software analytics. This occurs due to misuse of the term "software analytics" for describing traditional data mining, machine learning, or statistical work.
- Studies that were irrelevant to software projects, such as the automotive industry, that misuse the term "software analytics" to refer to general "data analytics."
- Studies that are relevant to generic data analytics and are not directly relevant to SA or software artifacts.

By applying both inclusion and exclusion criteria, the relevant papers numbered 41. After applying phase 2 of the filtration process, represented by the quality assessment stage (see the next section), the relevant papers were narrowed down to 19; these papers were used for data extraction. The list of selected studies is shown in Table I.

*D. Review Quality Assessment*

This step is important to ensure the accuracy of data extraction from the studies reviewed and in order to be confident about our results and conclusions. We defined the following quality assessment criteria:
1. QA1: The study contribution is clearly stated.
2. QA2: Software artifacts that are used are clearly explained.
3. QA3: SA characteristics are clear, different from those of direct statistics where advanced insights are provided.
4. QA4: The results and application(s) are described in detail.

Each of the quality assessment criteria has only three optional answers: "Yes" = 1, "Partly" = 0.5 and "No" =0. For each study, the quality score is the sum of the scores of each quality assessment point and the overall score is adjusted to a percentage scale. For our study, the quality assessment was used mainly as a selection criteria, as previously mentioned, based on the limitation that the papers considered are only those which have a quality score of $\geqslant 50\%$. The quality scores of the papers considered are shown in Table II.

*E. Data Extraction*

To obtain the data which is needed to address our research questions, we used the data extraction card shown in Table III.

*F. Data Synthesis*

In this stage, the extracted data was aggregated in order to answer the research questions. For our research questions, we used the narrative synthesis method. Accordingly, we will use tables and charts to present our results.

## III. RESULTS AND DISCUSSION

The dominant observation of our review was that there is not much relevant or mature research in the field of SA. This is clear from the number of papers considered (19) after applying both filtration phases as explained earlier. The number of publications shown included all studies that were available and reviewed. Results showed that about 79% of the considered papers (15) were from conferences while the remaining 21% (4) were from journals. Further, almost all journal papers (3) were from *IEEE software* and were included in SA special edition published in 2013. These statistics emphasize the fact of the difficulty we faced in finding mature SA work for our review. As mentioned in the quality assessment section, we considered only the papers with a quality score of ≥ 50% in order to guarantee including the most relevant studies. Most of the studies considered have a quality score of ≥ 75% (15 out of 19 papers). Table IV shows the quality score levels considering all papers that passed the first filtration phase.

The distribution of the studies selected in each publication year is shown in Fig.1, which clearly shows that SA studies became more active in the last two years, 2013 and 2014 only.

In the following subsections, we illustrate the review results for each of our research questions, one by one, supported with statistics from our data extraction.

*A. Beneficiary Practitioners (RQ1)*

***RQ1: Which software practitioners does the available SA research target?***

From the studies selected, we identified the main practitioners that the available SA studies support:
- Developer
- Tester

TABLE I. SELECTED PRIMARY STUDIES

| ID | Authors | Addressed Research Questions | | | | Ref. |
|---|---|---|---|---|---|---|
| S1 | M. van den Brand et al. | 1 | 2 | 3 | 4 | [7] |
| S2 | A. Gonzalez-Torres et al. | 1 | 2 | 3 | | [8] |
| S3 | E. Stroulia et al. | 1 | 2 | 3 | 4 | [9] |
| S4 | D. Reniers et al. | 1 | 2 | 3 | | [10] |
| S5 | R. Minelli and M. Lanza | 1 | 2 | 3 | 4 | [11] |
| S6 | J. Lou et al. | 1 | 2 | 3 | 4 | [12] |
| S7 | C. Klammer and J. Pichler | 1 | 2 | 3 | | [13] |
| S8 | T. Taipale et al. | 1 | 2 | 3 | 4 | [14] |
| S9 | O. Baysal et al. | 1 | 2 | 3 | | [15] |
| S10 | P. Johnson et al. | 1 | 2 | 3 | 4 | [16] |
| S11 | J. Czerwonka et al. | 1 | 2 | 3 | 4 | [17] |
| S12 | J. Gong and H. Zhang | 1 | 2 | 3 | 4 | [18] |
| S13 | A. Miranskyy et al. | 1 | 2 | 3 | 4 | [19] |
| S14 | R. Wu et al. | 1 | 2 | 3 | | [20] |
| S15 | S. Han et al. | 1 | 2 | 3 | | [21] |
| S16 | Y. Dubinsky et al. | 1 | 2 | 3 | | [22] |
| S17 | N. Chen et al. | 1 | 2 | 3 | | [23] |
| S18 | M. Mittal and A. Sureka | 1 | 2 | 3 | | [24] |
| S19 | G. Robles et al. | 1 | 2 | 3 | 4 | [25] |

TABLE II. QUALITY SCORES

| Study ID | QA1 | QA2 | QA3 | QA4 | Score |
|---|---|---|---|---|---|
| S1 | 1 | 1 | 0 | 1 | 75% |
| S2 | 1 | 1 | 0 | 1 | 75% |
| S3 | 1 | 1 | 1 | 1 | 100% |
| S4 | 1 | 1 | 0 | 1 | 75% |
| S5 | 1 | 1 | 0.5 | 0.5 | 75% |
| S6 | 1 | 1 | 1 | 0.5 | 87.5% |
| S7 | 0.5 | 1 | 0 | 0.5 | 50% |
| S8 | 1 | 1 | 1 | 0.5 | 87.5% |
| S9 | 1 | 1 | 0 | 1 | 75% |
| S10 | 1 | 1 | 1 | 0.5 | 87.5% |
| S11 | 1 | 1 | 1 | 1 | 100% |
| S12 | 1 | 1 | 1 | 1 | 100% |
| S13 | 1 | 1 | 1 | 1 | 100% |
| S14 | 0.5 | 1 | 0 | 0.5 | 50% |
| S15 | 1 | 1 | 0 | 0.5 | 62.5% |
| S16 | 0.5 | 1 | 0 | 0.5 | 50% |
| S17 | 1 | 1 | 0 | 1 | 75% |
| S18 | 1 | 1 | 0.5 | 1 | 87.5% |
| S19 | 1 | 1 | 1 | 1 | 100% |

TABLE III. THE DATA EXTRACTION CARD

| |
|---|
| Study id |
| Authors |
| Study title |
| Source |
| Year of publication |
| RQ1: Beneficiary practitioners |
| RQ2: Domain |
| RQ3: Analyzed software artifacts |
| RQ4: Different linked artifacts |

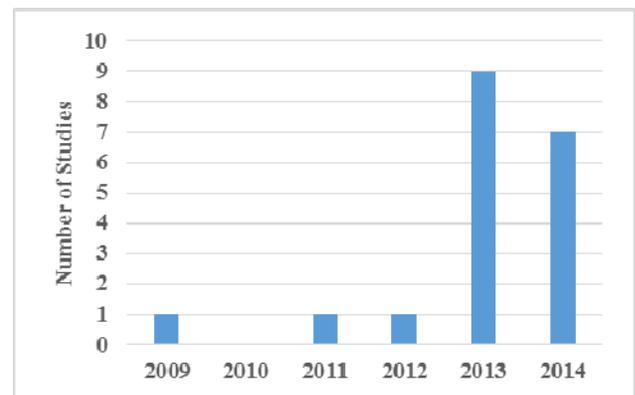

Fig. 1. Distribution of selected studies per year

TABLE IV. QUALITY ASSESSMENT LEVELS STATISTICS

| Quality Levels | # Studies | Percentage |
|---|---|---|
| Very high (85% ≤ score ≤ 100%) | 9 | 22% |
| High (75% ≤ score < 85%) | 6 | 15% |
| Medium (50% ≤ score < 75%) | 4 | 9% |
| Low (0% ≤ score < 50%) | 22 | 54% |

- Project Manager (PM)
- Portfolio Manager and High Management

As shown in Fig. 2, 90% of all studies targeted developers (17 out of 19) with about 47% (9) exclusively supporting developers (for details see Table V). These results show that SA needs more elaboration regarding stakeholders other than developers. Even available research work that supports other stakeholders, like PMs, is still immature and is similar to the direct statistics and dashboard work. For example, Stroulia et al. (S3) proposed a framework called "Collaboratorium Dashboard" in order to visualize insights extracted from collaborative software development tools that included information related to the team that has worked on a certain project, project artifacts, communication between project stakeholders, and the process followed. Also, the authors have integrated their framework with IBM Jazz and WikiDev, which already included integration with SVN, Bugzilla, email, and wikis. Although the proposed dashboard provided useful information for PMs in a visual form, such as the number of emails sent by each team member and the number of files checked in by each developer, this still formed a straight-forward insight extraction or statistics from software artifacts. More analytics are needed to link more than one artifact and get more supportive and powerful decisions. This can be the link between the source code of a certain feature, the emails related to this feature, or the quality reports in order to highlight the need for refactoring a certain part of this code. Such advanced analytics are a major need for any future research in SA.

*B. Research Domain (RQ2)*

***RQ2: Which domains are covered by SA studies?***

The aim of extracted data for RQ2 was the identification of the main active SA research domains in order to support practitioners in deciding both the hot topics and research opportunities. Our review showed that most available SA studies fell into one of the following domains:

- Maintainability and Reverse Engineering
- Team Collaboration and Dashboard
- Incident Management and Defect Prediction
- SA Platform
- Software Effort Estimation.

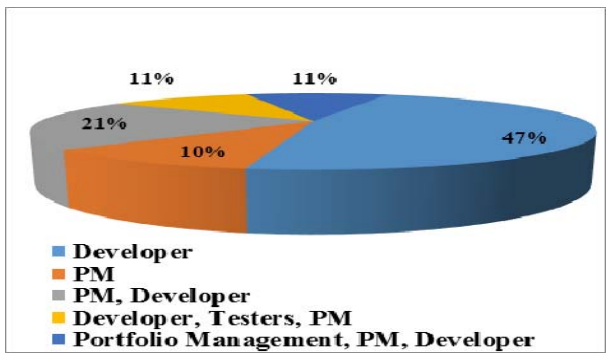

Fig. 2. Distribution of selected studies per practitioner

TABLE V. RQ1 EXTRACTED DATA

| Practitioner | Supporting Studies |
|---|---|
| Developer | S1, S4, S5, S6, S7, S8, S9, S10, S11, S13, S14, S15, S16, S17, S19 |
| Tester | S2, S13 |
| Project Manager | S2, S3, S4, S8, S10, S11, S12, S13, S18, S19 |
| Portfolio Manager | S10, S19 |

The distribution of the studies considered per domain can be found in Fig. 3 (for details see Table VI).

In the following subsections, we will illustrate our findings for the most significant studies in each domain.

*1) SA for Software Maintenance and Reverse Engineering*

Gonzalez-Torres et al. (S2) provided a visualization tool (Maleku) which extracts facts and insights from large legacy software and provides PMs and developers with useful information in order to support their decisions related to software maintenance. This tool extracts information from software repositories and monitors the repository for any updates in order to redo the analysis.

Although the proposed tool provided visualization features, these features simply represent traditional statistical information, like extracting the metrics related to inheritance and interface implementation.

Another study by Van den Brand et al. (S1), presents SQuAViSiT – a powerful visual software analytics tool. It has been successfully applied to the maintainability assessment of industry-sized software systems, combining analysis results of metrics (such as quality analysis), and visualization of these analysis results. The tool provides software design metrics such as cyclomatic complexity and inheritance depth. The tool also provides checking of code convention, duplication, and bad practices. Although the visual tool provided is useful, the introduced metrics analysis is traditional and appears in older

TABLE VI. RQ2 EXTRACTED DATA

| Domain | Studies |
|---|---|
| Maintainability and Reverse Engineering | S1, S2, S4, S5, S7, S12, S13, S14, S15, S16, S17 |
| Team Collaboration and Dashboard | S3, S9, S10, S18 |
| Incident Management and Defect Prediction | S6, S8 |
| Software Analytics Platform | S11 |
| Software Effort Estimation | S19 |

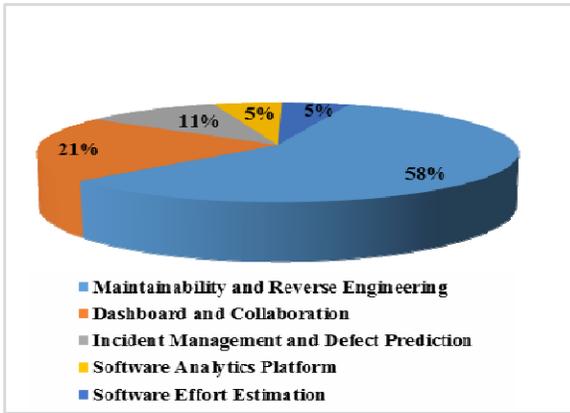

Fig. 3. Distribution of selected studies per domain

literature.

Minelli and Lanz (S5) are trying to figure out if the traditional maintainability approaches are valid for mobile applications (apps). They rely on the analytics of three artifacts: source code, third party API invocation, and application revision historical data. Minelli and Lanz implemented a visualization software analytics tool for mobile apps called "SAMOA" (Software Analytics for Mobile Applications). The tool provides visual presentation for multiple software metrics, apps versions, and the size of relative line of code between core functionality and third party invocation.

Although, the visualization tool presented can support project management, the metrics presented are very similar to traditional metrics from literature. It was expected to use more available artifacts such as user comments and ratings from app stores (like iOS Apple store or Google apps store). Also, Minelli and Lanz rely on only one dataset for their study.

Klammer and Pichler (S7) introduced a reverse engineering tool and applied it to electrical engineering software programs. The tool analyzes only the software source code in order to provide some insights related to source code structure and to locate features within source code. Multiple languages are supported such as C++ and Python. This work is similar to traditional work, and it needs to consider other software artifacts in order to apply software analytics concepts.

*2) SA for Team Collaboration and Dashboard*

Baysal et al. (S9) provided the Mozilla development team with a new qualitative dashboard as a complementary tool for the traditional quantitative reports of the Bugzilla issue tracking system. The qualitative dashboard provided improves development team awareness of the project situation and future directions. New features were provided, such as guiding developers to new information regarding their patches since the last check, highlighting new comments and reassigned patch reviewers. This research is promising since the trend toward qualitative analysis is powerful, and it can facilitate and speed up the decision-making process instead of relying on a deep statistical analysis as in traditional quantitative methodologies. However, the features provided are very direct and can be easily achieved by reviewing the bug history on the issue tracking system. In order to make this work more mature, new features such as team productivity trend charts can be provided.

*3) SA for Incident Management and Defect Prediction*

Lou et al. (S6) introduce a software analytics tool called Service Analysis Studio (SAS). SAS supports engineers in improving incident management by facilitating and automating the extraction of supportive insights. SAS has the ability to use multiple data sources – such as performance counters, operating system logs, and service transaction logs – to provide insights.

What makes this study important is that it applies the SA concept by linking multiple artifacts. Also, it presents a new algorithm to analyze system metrics data and suggests what abnormal metric is suspected of being the root cause of the incident. In addition, it introduces a mining technique to find the suspicious execution patterns, which are the sequence of actions that led to the incident, within the huge number of transaction logs.

*4) SA Platform*

Czerwonka et al. at Microsoft (S11) provide a software analytics common platform called CODEMINE. The need for CODEMINE came from the observation of the commonality between the input, outputs, and processing needs of multiple analytics team tools. CODEMINE acts as the common analytics framework for multiple client SA applications at Microsoft. The CODEMINE ability to provide data from different software artifacts (such as source code, project schedule, milestones, and defect reports) opens the opportunity for new research areas at Microsoft. In turn, this will enrich the insights by extracting information from cross-products which will boost team collaboration.

*5) SA for Software Effort Estimation*

G. Robles et al. (S19) present a study on the effort estimation of the *OpenStack* open-source project. Effort estimation of open source projects is challenging, as such projects have both a collaborative and distributed nature, and it is difficult to track the development effort. As a result, the authors offer a model that extracts data related to developer activities from the source code management repository and then guess the effort roughly based on these activities (like the time between two commits). Then the model calibrates the rough estimates based on other estimates collected from the developers in a survey. This study is promising, especially since it links artifacts to obtain insights that are useful to tackle such a hard-to-track topic as effort estimation of open source software projects.

*C. Analyzed Software Artifacts (RQ3)*

**RQ3: Which software artifacts are extracted?**

In order to address RQ3, we extracted the types of artifacts analyzed. This is very important for our study to evaluate the alignment of the studies with the goal of SA in analyzing more than one software artifact and providing more advanced insights.

The results of our review show that around 47% of the studies are still using only one artifact (9 studies), and many of these studies only analyze source code as in the case of traditional software analysis and metrics studies (4 studies). These results support our conclusion that most of the currently available SA studies are still immature and confused about the difference between the direct software analysis and the new SA. The results summary is shown in Fig. 4; more details can be found in Table VII.

*D. Checking Artifacts Linkage before Analysis (RQ4)*

***RQ4: If different artifacts are used, are they linked together?***

In order to address our last research question, RQ4, we evaluated the analysis of the artifacts used. Our main goal was to make sure that the artifacts were linked together in order to get more complex insights that could support software practitioners in making their decisions. It is worthwhile to highlight that this analysis was valid for only 10 studies when more than one artifact was used. This was achieved by reviewing the study scores for the third quality assessment criteria (QA3). The results showed that eight studies scored 100%. This shows that these studies link multiple artifacts to get insights that can support decision making. Therefore, these studies comply with the SA concept and can be considered as good references for practitioners to understand the SA concept. For more details, see quality scores in Table II.

## IV. LIMITATIONS OF THIS REVIEW

In our review, we considered both journal and conference papers without evaluating their rankings. This can be attributed to the difficulty that we faced when trying to find mature and relevant papers, and it was due to two reasons. The first reason is that the SA field is a new field of less than four years at the time of this review. The second reason is misuse of the term SA and the confusion of the researchers about its correct indication. This was shown by the number of papers considered after applying the filtration phases as previously mentioned.

## V. CONCLUSION AND FUTURE WORK

In this review, the available SA studies were investigated in order to understand the current research status of this new research topic. We conducted a literature review searching for the relevant studies available from 2000 - 2014. Our review considered 19 primary studies that supported us in addressing our four defined research questions. Our results can be summarized as follows:

- RQ1: The practitioners who benefit from the current SA studies are developers, testers, project managers (PM), portfolio managers, and higher management; about 47% of the considered studies supported only developers.
- RQ2: The studies considered showed that SA research covered the domains of maintainability and reverse engineering, team collaboration and dashboards, incident management and defect prediction, the SA platform, and software effort estimation.

TABLE VII. RQ3 EXTRACTED DATA

| Study ID | Analyzed Artifacts |
|---|---|
| S1, S4, S7, S16 | Source code |
| S2 | Code repository |
| S3 | Source code repository, issue tracking system, email, wikis |
| S5 | Source code, version control system |
| S6 | Performance counters, operating system logs, service transaction logs |
| S8 | Issue management system, version control system, code reviewing system, source code, organizational data, testing data |
| S9 | Issue tracker |
| S10 | Process data, product data |
| S11 | Source code, project schedule, milestones, defect reports |
| S12 | Source code, bug reports |
| S13 | Source code, version control system, bug reports |
| S14, S15 | Call stack |
| S17 | Mobile apps users reviews |
| S18 | Team wiki, version control system, issue tracking system |
| S19 | Version control system, developers survey |

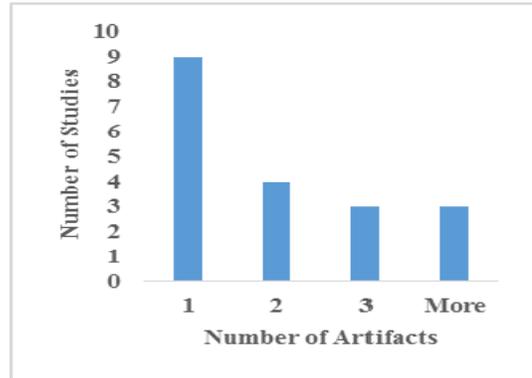

Fig. 4. Number of analyzed artifacts versus number of studies

- RQ3: Most of the studies considered (around 47%) are still analyzing only one artifact for their study.
- RQ4: Most of the studies considered analyze more than one artifact providing more complex insights, but there is still room for improvement of these studies. The review results showed that most of the available SA research introduces direct software statistics like design metrics and change history, simply decorated with some new analytics contributions such as linking team members to the classes they update. Also, most of the research addresses the low-level analytics of source code.

Based on our analysis, this review provides a recommendation for researchers that more research and elaboration need to be done, such as considering more artifacts in order to add value to traditional work and using more datasets to achieve higher confidence level in the results. In addition, there is a lack of research targeting higher-level business decision making like portfolio management, marketing strategy, and sales directions.

As future work, our study can be extended by considering more data sources such as *ScienceDirect*, *SpringerLink*, *Scopus*, *INSPEC*, following the references in the studies considered in this review, relevant journals and conferences. Also, we can widen our search term to include SA relevant terms such as mining software repositories.